\documentstyle[11pt]{article}
\begin{document}
\centerline{\Large\bf On Teleportation of a Completely Unknown State}
\vskip 3mm
\centerline{\Large\bf of Relativistic Photon}
\vskip 2mm
\centerline{R.Laiho$^{\dag}$, S.N.Molotkov$^{\ddag}$, and S.S.Nazin$^{\ddag}$}
\centerline{\it\small $^{\dag}$Wihuri Physical Laboratory,}
\centerline{\it\small Department of Physics,
University of Turku, 20014 Turku, Finland}
\centerline{\it\small $^{\ddag}$Institute of Solid State Physics of
                      Russian Academy of Sciences,}
\centerline{\it\small Chernogolovka, Moscow District, 142432 Russia}
\vskip 3mm

\begin{abstract}
The process of teleportation of a completely unknown single-photon
relativistic state is considered. Analysis of the relativistic case
reveals that the teleportation as it is understood in the non-relativistic
quantum mechanics is impossible if no {\it a priori} information on the
state to be teleported is available. It is only possible to speak of the
amplitude of the propagation of the field (taking into account
the measurement procedure) since the existence of a common
vacuum state together with the microcausality principle (the field
operators commutation relations) make the concept of the propagation
amplitude for the individual subsystems physically meaningless.
When partial {\it a priori} information is available (for example, only
the polarization state of the photon is unknown while its spatial
state is specified beforehand), the teleportation does become possible
in the relativistic case. In that case the {\it a priori} information
can be used to ``label'' the identical particles to make them
effectively distinguishable.
\end{abstract}
\vskip 3mm

One of the basic results of the non-relativistic quantum
information theory consists in the possibility of the ideal
teleportation of an unknown quantum state by means of a classical
and distributed quantum communication channel [1]. The latter
is realized by a non-local entangled state (an EPR-pair [2]).
The teleportation of an unknown quantum state from the user A to
user B is performed in the following way [1]. The user A has
a quantum system 1 in an unknown state $\rho_1$ (for example,
a spin-1/2 particle) whose state is to be teleported to user B.
To do this, the user A prepares a composite system consisting
of the two particles 2 and 3 in a maximally entangled state $\rho_{23}$
(EPR-pair) so that the user B can only access the particle 3 while
the particles 1 and 2 remain available to user A. Then user A
performs an appropriate joint measurement over the particles 1
(whose state is to be teleported) and 2. The measurement brings the
composite system consisting of the particles 1, 2, and 3 from
the initial state $\rho_1\otimes\rho_{23}$ to the new state $\rho^{'}_{123}$
which depends on the measurement outcome $i$ (in the teleportation
of the system described by a finite-dimensional state space the
set of all possible outcomes is discrete). The crucial point here
is the existence of the measurements possessing the special property
that the state of the third particle after the measurement
$\rho^{'}_{3}$ obtained by taking
a partial trace with respect to the states of particles 1 and 2,
$\rho^{'}_{3}=\mbox{Tr}_{12} \{ \rho^{'}_{123} \}$, coincides with the
unknown initial state $\rho_1$ to within a unitary transformation
which is completely determined by the measurement outcome $i$ obtained
by user A only:
\begin{equation}
\rho_1=U_i\rho^{'}_{3} U^{-1}_{i}.
\end{equation}
Here we identify the isomorphic state space of the subsystems 1 and 3.
The classical communication channel in the outlined teleportation
scheme is required to convey the measurement outcome $i$ from user A
to the distant user B. Preforming the measurement, the user A acquires
no information on the teleported state since all the measurement
outcomes are equiprobable.

Later, a number of different non-relativistic teleportation schemes
were proposed. All the schemes based on the non-relativistic
quantum mechanics employ the fact that the state space of a composite
system can be represented as the tensor product of the state spaces of
the constituent subsystems. In addition, it is important that the
measurement performed over a composite system can be represented
as a tensor product of the appropriate measurements over the
constituent subsystems.

The possibility of representation of both the state space and the
measurement as a tensor product stems from the following argument.
If one assumes that the composite system consists of physically distinct
(and hence distinguishable) particles, its state space can be written
as a tensor product of the corresponding states. Then the quantum
teleportation can be described using the standard mathematical
apparatus of quantum mechanics.

However, the non-relativistic quantum mechanics provides only an
approximate description of real world. A more complete description is
provided by the quantum field theory. In the quantum field theory
the superselection rules prohibit the superposition of pure states
belonging to different coherent sectors [3,4]. The analysis of
teleportation of identical particles belonging to the same coherent
sector in the quantum field theory proves to be substantially different
from the non-relativistic theory. First, the state space of a composite
system can no longer be considered as the tensor product of the
state spaces of constituent subsystems because of the existence of
the common cyclic vacuum vector [3,4]. Second, the microcausality
principle (commutation or anticommutation relations) requires that
the field operators commute (anticommute) if their supports are
separated by a space-like interval [3,4].

Using the photon field as an example, we shall show below that
the teleportation of the state of one of the identical particles
as it is understood in the non-relativistic quantum mechanics
is impossible in the quantum field theory because of the
existence of a common vacuum vector and impossibility of writing
a measurement over a composite system as a tensor product of
appropriate measurements over the constituent systems.
It is only possible to speak of the propagation amplitude of the field
as a whole.

It should be emphasized that our arguments do not formally prohibit
the possibility of teleportation of a completely unknown state in a
system of physically distinguishable particles (for example, a photon
state can be teleported to the electron degrees of freedom and
{\it vice versa}). However, a correct analysis of this question within
the framework of the quantum field theory for interacting fields
is much more difficult.

Consider now the teleportation of a completely unknown single-photon
state of the electromagnetic field.

The electromagnetic field operators can be written as [5]
\begin{equation}
A_{\mu}^{\pm}(\hat{x})=\frac{1}{(2\pi)^{3/2}}
\int \frac{d {\bf k}}{\sqrt{2k^0}}\mbox{e}^{\pm i\hat{k}\hat{x}}
e^{m}_{\mu}({\bf k)} a^{\pm}_{m}({\bf k}),
\end{equation}
\begin{displaymath}
\mu,m=1\ldots 4, \quad \hat{x}=(x^0,{\bf k}),\quad
\hat{k}=(k^0,{\bf k}),\quad k^0=|{\bf k}|,
\end{displaymath}
\begin{displaymath}
({\bf e}^m\cdot{\bf e}^n)=\delta_{m,n},\quad
(m,n=1\div 3),\quad
e_0^m=0,\quad
{\bf e}^3={\bf k}/|{\bf k}|.
\end{displaymath}
The four-dimensional vector potential operators satisfy the commutation
relations
\begin{equation}
[A_{\mu}^{-}(\hat{x}),A^{+}_{\nu}(\hat{x}')]=
ig_{\mu\nu}D^-_0(\hat{x}-\hat{x}'),
\end{equation}
where $g_{\mu\nu}$ ($g_00=-g_ii$, $i=1\div 3$), and
$D^-_0(\hat{x})$ is the commutator function for the massless field
\begin{equation}
D^{\pm}_{0}(\hat{x})=
\pm\frac{1}{i(2\pi)^{3/2}} \int \mbox{e}^{\pm i \hat{k}\hat{x}}
\theta(k^0)\delta(\hat{k}^2)d\hat{k}=
\pm\frac{1}{i(2\pi)^{3/2}}
\int \frac{d{\bf k}}{2k^0}\mbox{e}^{\pm i \hat{k}\hat{x}}=
\end{equation}
\begin{displaymath}
\frac{1}{4\pi}\frac{\delta(x^0-|{\bf x}|)-\delta(x^0+|{\bf x}|)}
{2|{\bf x}|}.
\end{displaymath}
There are for types of the creation operators ($a^{\pm}_{m}({\bf k})$) in
Eq. (2) describing two transverse, one temporal, and longitudinal photons
The last two photon types are unphysical and can be eliminated by introducing
an indefinite metrics [5]. For our purposes the shortest way to the
required result involves employment of a particular gauge. We shall
further work in the physical subspace of the two types of transverse
photons in the Coulomb gauge ($A_{\mu}=({\bf A},\varphi=0)$).

In that case the operator-valued distribution is a three-dimensional vector
\begin{equation}
\mbox{\boldmath $\psi$}(\hat{x})=
\frac{1}{(2\pi)^{3/2}}\sum_{s=\pm}\int \frac{d{\bf k}}{\sqrt{2k^0}}
{\bf w}({\bf k},s)
\{
a^-({\bf k},-s)\mbox{e}^{-i\hat{k}\hat{x}}+
a^+({\bf k},s)\mbox{e}^{i\hat{k}\hat{x}}
\},
\end{equation}
where ${\bf w}({\bf k},s)$ is a three-dimensional vector describing the
state helicity ($s=\pm$)
\begin{equation}
{\bf w}({\bf k},s)=\frac{1}{\sqrt{2}}({\bf e}_1({\bf k}) \pm i
{\bf e}_2({\bf k})),
\quad
{\bf e}_2({\bf k})\cdot {\bf e}_2({\bf k})=0,
\quad
{\bf e}_1({\bf k})\bot {\bf e}_2({\bf k})\bot {\bf k},
\end{equation}
where ${\bf e}_{1,2}({\bf k})$ are the linear polarization vectors.
The creation operators in Eq.(5) satisfy the commutation relations
\begin{equation}
[a^-({\bf k}\,s), a^+({\bf k},s')]=\delta({\bf k}-{\bf k}')\delta_{s,s'}.
\end{equation}
For convenience we introduce the new operators
\begin{equation}
{\bf A}^{\pm}({\bf k},s)= {\bf w}({\bf k},s)a^{\pm}({\bf k},s).
\end{equation}
The operator-valued distribution satisfies the Maxwell equation for
the free electromagnetic field
\begin{equation}
\nabla\times \mbox{\boldmath $\psi$}(\hat{x})=
-i\frac{\partial}{\partial x^0}\mbox{\boldmath $\psi$}(\hat{x}),
\end{equation}
\begin{displaymath}
\nabla\cdot \mbox{\boldmath $\psi$}(\hat{x})=0.\\
\end{displaymath}
Any state of the free electromagnetic field can be
obtained by acting with the field operators on the cyclic vacuum vector
\begin{equation}
|\mbox{\boldmath $\Psi$}\rangle=(
f_0+\sum_{n=1}^{\infty}\sum_{s_1=\pm}\sum_{s_2=\pm}\ldots\sum_{s_n=\pm}
\int\int\ldots\int
\frac{d{\bf k}_1}{\sqrt{2k^0_1}}
\frac{d{\bf k}_2}{\sqrt{2k^0_2}}
\ldots
\frac{d{\bf k}_n}{\sqrt{2k^0_n}}
\end{equation}
\begin{displaymath}
f({\bf k}_1,s_1,{\bf k}_2,s_2,\ldots{\bf k}_n,s_n,)
{\bf A}^{+}({\bf k}_1,s_1)
{\bf A}^{+}({\bf k}_2,s_2)
\ldots
{\bf A}^{+}({\bf k}_n,s_n)
)
|0\rangle,
\end{displaymath}
where $f({\bf k}_1,s_1,{\bf k}_2,s_2,\ldots{\bf k}_n,s_n,)$ is the amplitude
in the ${\bf k}$-representation.

The creation operator of the unknown single-particle field state
to be teleported is written as
\begin{equation}
\mbox{\boldmath $\psi$}^+({\bf f})=
\int \frac{d{\bf k}}{\sqrt{2k^0}}
\left( f({\bf x},+)\mbox{e}^{-ik^0x^0} {\bf A}^{+}({\bf k},+)+
f({\bf x},-)\mbox{e}^{-ik^0x^0} {\bf A}^{+}({\bf k},-) \right).
\end{equation}
It is convenient to write the temporal factor in Eq.(11) explicitly
although it can be incorporated in the definition of the single-particle
amplitude. In the position representation Eq.(11) becomes
\begin{equation}
\mbox{\boldmath $\psi$}^+({\bf f})=
\int d{\bf x}\left(
f({\bf x},+)\mbox{\boldmath $\psi$}^+(\hat{x},+)+
f({\bf x},-)\mbox{\boldmath $\psi$}^+(\hat{x},-)
\right).
\end{equation}
We have included the temporal factor in the argument of the field operator
in Eq.(12). Here $f({\bf x},\pm)$ is the Fourier transform of the amplitude
in the ${\bf k}$-representation. The quantity $f({\bf x},\pm)$ is interpreted
as the packet shape in the ${\bf x}$-representation (for helicities $\pm$)
at time $x^0$. Note that because of the transverse nature of the
electromagnetic field it is generally impossible to factorize
the polarization and spatial degrees of freedom (as it is frequently
done without any justifications in the description of experiments when only
the two-dimensional polarization states space is considered).

Let us now construct the entangled EPR-state. The relativistic counterpart
of a maximally entangled EPR-state with respect to the polarization
(and, because of the transverse nature of the field, automatically with
respect to the momentum) is
\begin{equation}
|\mbox{\boldmath $\Psi$}_{00}\rangle=
\left(
\int \frac{d{\bf k}}{2k^0}\mbox{e}^{-2ik^0x^0}
{\bf A}^+({\bf k},+) {\bf A}^+({\bf -k},-)
\right)|0\rangle=
\left(
\int d{\bf x}
\mbox{\boldmath $\psi$}^+(\hat{x},+)
\mbox{\boldmath $\psi$}^+(\hat{x},-)
\right)
|0\rangle,
\end{equation}
where for convenience the temporal factor is again introduced which is later
included into field operator argument in Eq.(13). The EPR-state formally
correspond to the two-particle amplitude chosen in the form
${\cal F}(\hat{x}_1,s_1,\hat{x}_2,-s_2)= 
\delta_{s_1,-s_2}                        
\delta(x_1^0-x^0)                        
\delta(x_2^0-x^0)                        
\delta({\bf x}_1-{\bf x}_2)              
const({\bf x}_1+{\bf x}_2)$.
The latter should be understood as a limit of smooth (test) functions.
Formally, the EPR state (13) can be interpreted as describing
the creation of two photons with $s_1=-s_2$ at time $x^0$ at the point
${\bf x}_1={\bf x}_2$ in a correlated non-local way simultaneously in the
entire space as suggested by the factor $const({\bf x}_1+{\bf x}_2)$.

The EPR-state (13) corresponds to a pair of photons with zero (in the
chosen reference frame) total momentum. Since the polarization degrees
of freedoms generally do not factor out from the spatial ones, there
exist a continuum of maximally entangled EPR-states with different
total momentum. Although any of these states can equally well be
used for our purposes, we shall use the pair with zero total momentum
to make the analogy with the non-relativistic Bell basis more clear.

The states analogous to the non-relativistic Bell basis are
\begin{equation}
|\mbox{\boldmath $\Psi$}^{(\pm)}_{\bf YP}\rangle=
\mbox{\boldmath $\Psi$}^{(\pm)+}_{\bf YP}|0\rangle=
\frac{1}{\sqrt{2}}
\left(
\int d\mbox{\boldmath $\xi$}
\mbox{e}^{-i\mbox{\boldmath $\xi$}{\bf P}}
(
\mbox{\boldmath $\psi$}^+(\hat{\xi},+)
\mbox{\boldmath $\psi$}^+(\hat{\xi}-{\bf Y},-)\pm
\mbox{\boldmath $\psi$}^+(\hat{\xi},-)
\mbox{\boldmath $\psi$}^+(\hat{\xi}-{\bf Y},+)
)
\right)
|0\rangle,
\end{equation}
\begin{displaymath}
|\mbox{\boldmath $\Phi$}^{(\pm)}_{\bf YP}\rangle=
\mbox{\boldmath $\Phi$}^{(\pm)+}_{\bf YP}|0\rangle=
\frac{1}{\sqrt{2}}
\left(
\int d\mbox{\boldmath $\xi$}
\mbox{e}^{-i\mbox{\boldmath $\xi$}{\bf P}}
(
\mbox{\boldmath $\psi$}^+(\hat{\xi},+)
\mbox{\boldmath $\psi$}^+(\hat{\xi}-{\bf Y},+)\pm
\mbox{\boldmath $\psi$}^+(\hat{\xi},-)
\mbox{\boldmath $\psi$}^+(\hat{\xi}-{\bf Y},-)
)
\right)
|0\rangle.
\end{displaymath}
It is easily checked that these states constitute an orthogonal
identity resolution in the subspace of two-particle states. Indeed,
\begin{equation}
I_2=\int\int
\frac{d{\bf Y}d{\bf P}}{(2\pi)^3}
\end{equation}
\begin{displaymath}
\left(
|\mbox{\boldmath $\Psi$}^{(+)}_{\bf YP}\rangle
\langle\mbox{\boldmath $\Psi$}^{(+)}_{\bf YP}|+
|\mbox{\boldmath $\Psi$}^{(-)}_{\bf YP}\rangle
\langle\mbox{\boldmath $\Psi$}^{(-)}_{\bf YP}|+
|\mbox{\boldmath $\Phi$}^{(+)}_{\bf YP}\rangle
\langle\mbox{\boldmath $\Phi$}^{(+)}_{\bf YP}|+
|\mbox{\boldmath $\Phi$}^{(-)}_{\bf YP}\rangle
\langle\mbox{\boldmath $\Phi$}^{(-)}_{\bf YP}|
\right)=
\end{displaymath}
\begin{displaymath}
\sum_{s_1,s_2=\pm}
\int \int \frac{d {\bf k}_1}{2k^0_2} \frac{d {\bf k}_2}{2k^0_2}
\left(
{\bf A}^+({\bf k}_1,s_1) {\bf A}^+({\bf k}_2,s_2)|0\rangle
\right)
\left(
\langle 0|{\bf A}^-({\bf k}_1,s_1) {\bf A}^-({\bf k}_2,s_2)
\right)
\end{displaymath}
It should be emphasized once again that because of the transverse nature
of the electromagnetic field it is impossible to construct an identity
resolution (as well as an EPR-pair) with factorized polarization and spatial
degrees of freedom.

The initial common (because of the common vacuum vector) state
of the field describing the completely unknown state to be teleported
and the EPR-pair is written as
\begin{equation}
|\mbox{\boldmath $\psi$}({\bf f})
\mbox{\boldmath $\Psi$}^{(-)}_{\bf YP}\rangle=
\mbox{\boldmath $\psi$}^{+}({\bf f})
\mbox{\boldmath $\Psi$}^{(-)+}_{\bf YP}|0\rangle=
\int\int d{\bf x} d\mbox{\boldmath $\xi$}
\mbox{e}^{-i\mbox{\boldmath $\xi$}{\bf P}}
\end{equation}
\begin{displaymath}
\left\{
-
\mbox{\boldmath $\Psi$}^{(-)+}(\hat{\xi},\hat{x})
(
f({\bf x},+) \mbox{\boldmath $\psi$}^+(\hat{\xi}-{\bf Y},+)+
f({\bf x},-) \mbox{\boldmath $\psi$}^+(\hat{\xi}-{\bf Y},-)
)+
\right.
\end{displaymath}
\begin{displaymath}
\left.
\mbox{\boldmath $\Psi$}^{(+)+}(\hat{\xi},\hat{x})
(
-f({\bf x},+) \mbox{\boldmath $\psi$}^+(\hat{\xi}-{\bf Y},+)+
f({\bf x},-) \mbox{\boldmath $\psi$}^+(\hat{\xi}-{\bf Y},-)
)+
\right.
\end{displaymath}
\begin{displaymath}
\left.
\mbox{\boldmath $\Phi$}^{(-)+}(\hat{\xi},\hat{x})
(
f({\bf x},+) \mbox{\boldmath $\psi$}^+(\hat{\xi}-{\bf Y},-)+
f({\bf x},-) \mbox{\boldmath $\psi$}^+(\hat{\xi}-{\bf Y},+)
)+
\right.
\end{displaymath}
\begin{displaymath}
\left.
\mbox{\boldmath $\Phi$}^{(+)+}(\hat{\xi},\hat{x})
(
f({\bf x},+) \mbox{\boldmath $\psi$}^+(\hat{\xi}-{\bf Y},-)-
f({\bf x},-) \mbox{\boldmath $\psi$}^+(\hat{\xi}-{\bf Y},+)
)
\right\}
|0\rangle.
\end{displaymath}
Here the following notation is introduced:
\begin{equation}
\mbox{\boldmath $\Psi$}^{(\pm)+}(\hat{\xi},\hat{x})=
\frac{1}{\sqrt{2}}
\left(
\mbox{\boldmath $\psi$}^+(\hat{x},+)\mbox{\boldmath $\psi$}^+(\hat{\xi},-)
\pm \mbox{\boldmath $\psi$}^+(\hat{x},-)\mbox{\boldmath $\psi$}^+(\hat{\xi},+)
\right),
\end{equation}
\begin{displaymath}
\mbox{\boldmath $\Phi$}^{(\pm)+}(\hat{\xi},\hat{x})=
\frac{1}{\sqrt{2}}
\left(
\mbox{\boldmath $\psi$}^+(\hat{x},+)\mbox{\boldmath $\psi$}^+(\hat{\xi},+)
\pm
\mbox{\boldmath $\psi$}^+(\hat{x},-)\mbox{\boldmath $\psi$}^+(\hat{\xi},-)
\right),
\end{displaymath}
The temporal factors are again included in the arguments $\hat{x}$ and
$\hat{\xi}$. The identical transformations in Eq. (16) are performed to
make the analogy with the non-relativistic case [1] more graphical.

Let us now discuss the construction of the appropriate measurement.
We shall go back for a moment to the non-relativistic case. The change
of the state of a quantum system after a measurement act is completely
described by the corresponding instrument (superoperator) [5,6].
In the context of the teleportation problem, when the measurement affects
only the particle in the unknown state to be teleported (particle 1) and one
of the particles in the EPR-pair (particle 2) while the particle 3 is not
directly affected by the measurement, such an instrument with the space
of all possible outcomes $\Theta$ is written as
$\mbox{\bf T}_{123}(d\theta)=\mbox{\bf T}_{12}(d\theta)\otimes I_3$,
where $I_3$ is the identity operator in the state space of the third particle.
The system state just after the measurement which gave the outcome in the
interval $(\theta,\theta+d\theta)$ is
\begin{equation}
\rho^{'}_{123}=\frac{\textstyle \mbox{\bf T}_{123}(d\theta)\circ\rho_{1233} }
{\mbox{Tr}_{123} \{ \mbox{\bf T}_{123}(d\theta)\circ\rho_{123} \} }
\end{equation}
The density matrix of particle 3 after the measurement is obtained
by taking the partial trace over the states of particles 1 and 2:
\begin{equation}
\rho^{'}_{3}=
\frac{\textstyle \mbox{ Tr}_{12}\{
\mbox{\bf T}_{123}(d\theta)\circ\rho_{123}\} }
{\mbox{Tr}_{123} \{ \mbox{\bf T}_{123}(d\theta)\circ\rho_{123} \} }.
\end{equation}
If the composite system S consists of two subsystems A and B with the
density matrix $\rho_{AB}$ (for the teleportation problem, the subsystem
A consists of particles 1 and 2 while subsystem B consists of particle 3),
the state of subsystem B just after the measurement is (to within
the normalization constant)
\begin{equation}
\rho^{'}_{B}=
\mbox{ Tr}_{A}\{ \mbox{\bf T}_{AB}(d\theta)\circ\rho_{AB}\} =
\mbox{ Tr}_{A}\{ ({\cal M}_{A}(d\theta)\otimes I_B)\rho_{AB}\};
\end{equation}
here ${\cal M}_{A}(d\theta)=[\mbox{\bf T}_{AB}(d\theta)]^*I_{AB}$ is
a positive operator-valued measure on the space of possible outcomes
$\Theta$ generated by the instrument. Therefore, in the teleportation
problem it is sufficient to know only the measurement
${\cal M}_{A}(d\theta)$ [7] rather than the instrument itself
(generally, to determine the system state just after the measurement one
should know the corresponding instrument itself).
However, this statement is only valid if the instrument (and measurement)
can be represented as a tensor product of the instruments (measurements)
acting in the state spaces of the subsystems.

In the quantum field theory the measurement cannot be in principle
represented as a tensor product of the measurements related to
the individual subsystems because of the following two reasons.
The first one is the existence of the common vacuum vector. The second
reason is the microcausality principle (commutation relation given by Eq. (2)).
The possibility of the representation of the measurement over identical
particles in the form of a tensor product implies
that the field operators related to two different factors (particles)
always commute irrespective of the relative position of their supports
in the Minkowski space. Hence in the quantum field theory the measurement
over a composite system consisting of three particles can only be written
as a general identity resolution in the entire space of three-particle states.
However, then the teleportation itself as it is understood in non-relativistic
quantum mechanics becomes meaningless and one can only speak of the
amplitude of propagation of the field as a whole rather than the
individual subsystems.

Nevertheless, it is still interesting to find out at the level of
concrete formulas where the two indicated circumstances come into play.

We shall further need the following identity resolution in the one-particle
state space:
\begin{equation}
I_1=\sum_{s=\pm}\int\frac{d{\bf k}}{2k^0}
\left({\bf A}^+({\bf k},s)|0\rangle\right)
\left(\langle 0|{\bf A}^-({\bf k},s)\right)=
\end{equation}
\begin{displaymath}
\sum_{s=\pm}\int d{\bf x}
\left(\mbox{\boldmath $\psi$}^{+}({\bf x},s)|0\rangle\right)
\left(\langle 0|\mbox{\boldmath $\psi$}^{-}({\bf x},s)\right).
\end{displaymath}
The set of possible measurement outcomes $\Theta$ for the two-particle
state space is
$\Theta=$ $\{i,k,s,{\bf X},{\bf Q}:                   
i\times k\times(\pm)\times{\bf R}_{\bf X}\times{\bf R}_{\bf Q}\}$,
where $i=1,2$ and $k=\pm$ label the states of the Bell basis.
The total identity resolution in the three-particle state space is
\begin{equation}
I_3=\sum_{s=\pm,k=\pm,i=1,2}\int
{\cal M}^{k}_{is}(d\theta),
\end{equation}
where
\begin{displaymath}
{\cal M}^{\pm}_{1s}(d\theta)=
\left(
\mbox{\boldmath $\psi$}^+(\hat{x},s)
|\mbox{\boldmath $\Psi$}^{(\pm)}_{\bf XQ}\rangle
\right)
\left(
\langle \mbox{\boldmath $\Psi$}^{(\pm)}_{\bf XQ}|
\mbox{\boldmath $\psi$}^-(\hat{x},s)
\right)
\frac{d{\bf x}d{\bf X}d{\bf Q}}{(2\pi)^3}
,
\end{displaymath}
\begin{displaymath}
{\cal M}^{\pm}_{2s}(d\theta)=
\left(
\mbox{\boldmath $\psi$}^+(\hat{x},s)
|\mbox{\boldmath $\Phi$}^{(\pm)}_{\bf XQ}\rangle
\right)
\left(
\langle \mbox{\boldmath $\Phi$}^{(\pm)}_{\bf XQ}|
\mbox{\boldmath $\psi$}^-(\hat{x},s)
\right)
\frac{d{\bf x}d{\bf X}d{\bf Q}}{(2\pi)^3}
.
\end{displaymath}
The resolution (22) is an analogue of the identity resolution of the form
${\cal M}_{12}\otimes I_1$ arising in the non-relativistic analysis
although Eq.(22) does not reduce to the latter because of the two indicated
reasons. It should be noted that the same time $x^0$ appears in all arguments
in Eq. (22). Such a measurement can be interpreted as a spatially non-local
measurement performed at time $x^0$.

Note also that unlike the orthogonal identity resolution (15)
in the two-particle state space the resolution in the three-particle
space (in contrast to the non-relativistic case) becomes non-orthogonal.
Of course, this measurement can be made orthogonal by choosing
the symmetry-adapted basis functions realizing the irreducible
representations of the permutation group in the three-particle space,
but it cannot be made orthogonal when restricted to the two-particle space.

The probabilities of various measurement outcomes from the space $\Theta$ are
given by the standard formula
\begin{equation}
\mbox{Pr}\{s,i,k,d\theta \}=\mbox{Tr}\{
|\mbox{\boldmath $\psi$}({\bf f})
\mbox{\boldmath $\Psi$}^{(-)}_{\bf YP}\rangle
\langle \mbox{\boldmath $\Psi$}^{(-)}_{\bf YP}\rangle
\mbox{\boldmath $\psi$}({\bf f})|
{\cal M}^{k}_{is}(d\theta)
\}=
|{\cal A}_{is}^{k}(d\theta)|^2,
\end{equation}
where the amplitude ${\cal A}_{is}^{k}(d\theta)$ is defined, for example,
in the channel ${\cal M}^{-}_{s1}(d\theta)$ as
\begin{equation}
{\cal A}_{1s}^{-}(d\theta)=
\langle 0|
\mbox{\boldmath $\Psi$}^{(-)-}_{\bf XQ}
\mbox{\boldmath $\psi$}^{-}(\hat{x},s)
\mbox{\boldmath $\psi$}^{+}({\bf f})
\mbox{\boldmath $\Psi$}^{(-)+}_{\bf YP}
|0\rangle d\theta=
\end{equation}
\begin{displaymath}
\frac{1}{2}\int\int\int d{\bf x}' d\mbox{\boldmath $\xi$}'
d\mbox{\boldmath $\xi$}
\mbox{e}^{-i(\mbox{\boldmath $\xi$}{\bf P}- \mbox{\boldmath $\xi$}'{\bf Q})}
\end{displaymath}
\begin{displaymath}
\left\{
\langle 0|-
\mbox{\boldmath $\Psi$}^{(-)-}_{\bf XQ}
\mbox{\boldmath $\psi$}^{-}(\hat{x},s)
\mbox{\boldmath $\Psi$}^{(-)+}(\hat{\xi},\hat{x}')
\left(
f({\bf x}',+)\mbox{\boldmath $\psi$}^{+}(\hat{\xi}-{\bf Y},+)+
f({\bf x}',-)\mbox{\boldmath $\psi$}^{+}(\hat{\xi}-{\bf Y},-)
\right)
|0\rangle
+
\right.
\end{displaymath}
\begin{displaymath}
\langle 0|
\mbox{\boldmath $\Psi$}^{(-)-}_{\bf XQ}
\mbox{\boldmath $\psi$}^{-}(\hat{x},s)
\mbox{\boldmath $\Psi$}^{(-)+}(\hat{\xi},\hat{x}')
\left(
-f({\bf x}',+)\mbox{\boldmath $\psi$}^{+}(\hat{\xi}-{\bf Y},+)+
f({\bf x}',-)\mbox{\boldmath $\psi$}^{+}(\hat{\xi}-{\bf Y},-)
\right)
|0\rangle
+
\end{displaymath}
\begin{displaymath}
\langle 0|
\mbox{\boldmath $\Psi$}^{(-)-}_{\bf XQ}
\mbox{\boldmath $\psi$}^{-}(\hat{x},s)
\mbox{\boldmath $\Phi$}^{(-)+}(\hat{\xi},\hat{x}')
\left(
-f({\bf x}',+)\mbox{\boldmath $\psi$}^{+}(\hat{\xi}-{\bf Y},-)+
f({\bf x}',-)\mbox{\boldmath $\psi$}^{+}(\hat{\xi}-{\bf Y},+)
\right)
|0\rangle
+
\end{displaymath}
\begin{displaymath}
\left.
\langle 0|
\mbox{\boldmath $\Psi$}^{(-)-}_{\bf XQ}
\mbox{\boldmath $\psi$}^{-}(\hat{x},s)
\mbox{\boldmath $\Phi$}^{(+)+}(\hat{\xi},\hat{x}')
\left(
f({\bf x}',+)\mbox{\boldmath $\psi$}^{+}(\hat{\xi}-{\bf Y},-)-
f({\bf x}',-)\mbox{\boldmath $\psi$}^{+}(\hat{\xi}-{\bf Y},+)
\right)
|0\rangle
\right\}
d\theta
.
\end{displaymath}
It is impossible to identify the individual contribution of different
subsystems to the field propagation amplitude. Nevertheless, it is
interesting to consider the transition to the non-relativistic limit.
In that case, if one neglects the commutation relations and assumes
that the operators $\mbox{\boldmath $\Psi$}^{(\pm)\pm}_{\bf XQ}$
($\mbox{\boldmath $\Phi$}^{(\pm)\pm}_{\bf XQ}$),
$\mbox{\boldmath $\psi$}^{-}(\hat{x},s)$ and
$\left(f({\bf x}',+)\mbox{\boldmath $\psi$}^{+}(\hat{\xi}-{\bf Y},+)+ 
f({\bf x}',-)\mbox{\boldmath $\psi$}^{+}(\hat{\xi}-{\bf Y},-)
\right)$, etc. commute then, because of the orthogonality of different
$\mbox{\boldmath $\Psi$}^{(\pm)\pm}_{\bf XQ}$  and
$\mbox{\boldmath $\Phi$}^{(\pm)\pm}_{\bf XQ}$, only one term is left
in Eq. (24), for example, only the outcome related to the projection
on $\mbox{\boldmath $\Psi$}^{(-)-}_{\bf XQ}$ in which case we obtain
\begin{equation}
{\cal A}_{1s}^{-}(d\theta)\rightarrow
-\frac{1}{2}\int\int\int d{\bf x}' d\mbox{\boldmath $\xi$}'
d\mbox{\boldmath $\xi$}
\mbox{e}^{-i(\mbox{\boldmath $\xi$}{\bf P}- \mbox{\boldmath $\xi$}'{\bf Q})}
\end{equation}
\begin{displaymath}
\langle 0|
\mbox{\boldmath $\Psi$}^{(-)-}_{\bf XQ}
\mbox{\boldmath $\psi$}^{-}(\hat{x},s)
\mbox{\boldmath $\Psi$}^{(-)+}(\hat{\xi},\hat{x}')
\left(
f({\bf x}',+)\mbox{\boldmath $\psi$}^{+}(\hat{\xi}-{\bf Y},+)+
f({\bf x}',-)\mbox{\boldmath $\psi$}^{+}(\hat{\xi}-{\bf Y},-)
\right)
|0\rangle
d\theta
.
\end{displaymath}
For the free field the vacuum average in Eq. (25) factorizes into the
pair-wise averages. Assuming again that
$\mbox{\boldmath $\Psi$}^{(-)-}_{\bf XQ}$ and
$\mbox{\boldmath $\Psi$}^{(-)+}(\hat{\xi},\hat{x}')$ are related to
the particles 1 and 2 (particle in the unknown state and one of the EPR-pair
particles) and act in the appropriate subsystem state space while the
operators $\mbox{\boldmath $\psi$}^{-}(\hat{x},s)$ and
$\left(f({\bf x}',+)\mbox{\boldmath $\psi$}^{+}(\hat{\xi}-{\bf Y},+)+ 
f({\bf x}',-)\mbox{\boldmath $\psi$}^{+}(\hat{\xi}-{\bf Y},-)\right)$
act in the third particle state space (the second photon
in the EPR-pair), one obtains
\begin{equation}
{\cal A}_{1s}^{-}(d\theta)\rightarrow
-\frac{1}{2}\int\int\int d{\bf x}' d\mbox{\boldmath $\xi$}'
d\mbox{\boldmath $\xi$}
\mbox{e}^{-i(\mbox{\boldmath $\xi$}{\bf P}- \mbox{\boldmath $\xi$}'{\bf Q})}
\langle 0|
\mbox{\boldmath $\Psi$}^{(-)-}_{\bf XQ}
\mbox{\boldmath $\Psi$}^{(-)+}(\hat{\xi},\hat{x}')
|0\rangle
\end{equation}
\begin{displaymath}
\langle 0|
\mbox{\boldmath $\psi$}^{-}(\hat{x},s)
\left(
f({\bf x}',+)\mbox{\boldmath $\psi$}^{+}(\hat{\xi}-{\bf Y},+)+
f({\bf x}',-)\mbox{\boldmath $\psi$}^{+}(\hat{\xi}-{\bf Y},-)
\right)
|0\rangle.
\end{displaymath}
The amplitude in Eq. (26) has a rather transparent interpretation.
The first factor describes the propagation of a ``new'' EPR-pair
after the measurement affecting the particle 1 (which was initially
in a completely unknown state) and one of the particles of the
original EPR-pair. This is described by the operator
$\mbox{\boldmath $\Psi$}^{(-)+}(\hat{\xi},\hat{x}')$.
Corresponding to the measurement in the Bell basis is the action of
the operator of annihilation of the new EPR-pair
$\mbox{\boldmath $\Psi$}^{(-)-}_{\bf XQ}$.

The second factor describes the propagation of the second particle
from the original EPR-pair which due to the initial correlations in
the EPR-pair is brought after the performed Bell measurement into
the state
($\left(                                                         
f({\bf x}',+)\mbox{\boldmath $\psi$}^{+}(\hat{\xi}-{\bf Y},+)+   
f({\bf x}',-)\mbox{\boldmath $\psi$}^{+}(\hat{\xi}-{\bf Y},-)    
\right)$ which coincides to within a unitary transformation
completely determined by the measurement outcome with the unknown
state of the particle 1 which had to be teleported.
The fact that the measurement did not affect the particle 3 corresponds
to the annihilation operator $\mbox{\boldmath $\psi$}^{-}(\hat{x},s)$
formally describing the projection onto all one-particle states
(because of the integration over ${\bf x}$ in the expression (23)
for the probability).

One can advance further in this way using the explicit expression
for the first factor in Eq. (26) which gives
\begin{equation}
\langle 0|
\mbox{\boldmath $\Psi$}^{(-)-}_{\bf XQ}
\mbox{\boldmath $\Psi$}^{(-)+}(\hat{\xi},\hat{x}')
|0\rangle=
2\left(
{\cal D}^+_0(\hat{\xi}'-\hat{x}'){\cal D}^+_0(\hat{\xi}'-\hat{\xi}-{\bf X})-
{\cal D}^+_0(\hat{\xi}'-\hat{\xi}){\cal D}^+_0(\hat{\xi}'-\hat{x}'-{\bf X})
\right).
\end{equation}
The commutator function ${\cal D}^+_0(\hat{x}-\hat{y})$
describes the process of particle creation at point $\hat{x}$, its
propagation, and annihilation at point $\hat{y}$ (for $y^0>x^0$) [5]:
\begin{equation}
\langle 0 |
\mbox{\boldmath $\psi$}^{-}(\hat{y},s),
\mbox{\boldmath $\psi$}^{+}(\hat{x},s)
| 0 \rangle =
-i {\cal D}^+_0(\hat{x}-\hat{y}).
\end{equation}
The second term in Eq. (27) arises because of the particles
indistinguishability (propagation with the exchange of the particles)
If one again assumes that the creation and annihilation operators
related to the particles 1 and 2 act in the different state spaces
(which would correspond to the distinguishability of the particles),
then the second term should be discarded.

Finally, to obtain the non-relativistic limit, the commutator
functions ${\cal D}^+_0(\hat{x})$ in Eq. (27) should be replaced
by the ordinary $\delta$-functions of the three-dimensional argument
($\delta({\bf x})$). This replacement of ${\cal D}^+_0( \hat{x} )$-functions
by ordinary $\delta({\bf x})$-functions should be done because in the
non-relativistic case the integration is performed with a Galilei-invariant
measure $d\mu({\bf k})=d{\bf k}$ while in the relativistic case one employs
the Lorentz-invariant measure
$d\mu({\bf k})=\theta(k^0)\delta(\hat{k}^2)d\hat{k}=d{\bf k}/2k^0$
which finally yields
\begin{equation}
{\cal D}^+_0(\hat{x})= -\frac{1}{i(2\pi)^{3/2}} \int
\frac{d{\bf k}}{2k_0} \mbox{e}^{i\hat{k}\hat{x}} \rightarrow
\frac{1}{(2\pi)^{3/2}}
\int d{\bf k}
\mbox{e}^{ i{\bf kx} }=\delta({\bf x}).
\end{equation}
Then we obtain the final expression for the amplitude of the
teleported state (in the non-relativistic case the amplitude actually
coincides with the wave function)
\begin{equation}
{\cal A}_{1s}^{-}(d\theta)\rightarrow
-\frac{1}{2}\int\int\int d{\bf x}' d\mbox{\boldmath $\xi$}'
d\mbox{\boldmath $\xi$}
\mbox{e}^{-i(\mbox{\boldmath $\xi$}{\bf P}- \mbox{\boldmath $\xi$}'{\bf Q})}
\delta(\mbox{\boldmath $\xi$}-{\bf x}')
\delta(\mbox{\boldmath $\xi$}'-\mbox{\boldmath $\xi$}-{\bf X})
\end{equation}
\begin{displaymath}
\delta({\bf x}-\mbox{\boldmath $\xi$}-{\bf Y})
\left( \delta_{s,+} f({\bf x}^{'},+)+ \delta_{s,-}f({\bf x}^{'},-) \right)
=
\end{displaymath}
\begin{displaymath}
\left(
\delta_{s,+}f({\bf x}-{\bf Y},+)+
\delta_{s,-}f({\bf x}-{\bf Y},-)
\right)
\mbox{e}^{-i({\bf x}-{\bf Y})({\bf P}-{\bf Q})+i{\bf XQ}},
\end{displaymath}
which coincides (to within a unitary transformation) with the
original completely unknown state to be teleported.

{\it Thus, if the photon state to be teleported is completely unknown,
the teleportation as it is understood in the non-relativistic quantum
mechanics of distinguishable particles becomes impossible and one
can only speak of the amplitude of the propagation
(taking into account the measurement procedure) as whole.}

However, if the state to be teleported is only partly unknown,
and one has only to teleport several degrees of freedom
(for example, only the polarization state), the rest degrees of freedom
can be used to ``label'' the individual particles to make them effectively
distinguishable in the teleportation.

To be more precise, we mean the following. Suppose that it is known
in advance that the single-photon state to be teleported has a specified
momentum ${\bf k}_1$ and only the polarization state is unknown.
Such a state can be written as
\begin{equation}
\mbox{\boldmath $\psi$}^+({\bf f})=
\frac{1}{\sqrt{2k^0_1}}
\left(
f(+){\bf A}^+({\bf k}_1,+)  +  f(-){\bf A}^+({\bf k}_1,-)
\right),
\end{equation}
\begin{displaymath}
|f(+)|^2+|f(-)|^2=1,
\end{displaymath}
where $f(\pm)$ are the amplitudes for different polarizations $\pm$.
The phase factors are omitted as unimportant. A plane wave
is infinitely extended in space and it would be more correct
to consider a ray-like state. However, for our purposes the difference
between these states is insignificant and we shall use the simplest plane
wave state.

The EPR-state entangled with respect to the polarization degrees of freedom
is described by the creation operator
\begin{equation}
\mbox{\boldmath $\Psi$}^{(-)+}({\bf k}_2,{\bf k}_3)=
\frac{1}{\sqrt{2}\sqrt{2k^0_2}\sqrt{2k^0_3}}
\left(
{\bf A}^+({\bf k}_2,+){\bf A}^+({\bf k}_3,-)  -
{\bf A}^+({\bf k}_2,-){\bf A}^+({\bf k}_3,+)
\right),
\end{equation}
where the wave vectors ${\bf k}_2$ and ${\bf k}_3$ are also
known beforehand. In the experiments this state is associated with
a pair of photons in an entangled state leaving a non-linear crystal
and propagating along the vectors ${\bf k}_2$ and ${\bf k}_3$ [8].

The required measurement is described by the identity resolution (22).
It is more convenient here to rewrite it in the momentum representation
choosing
$\Theta=$
$\{s,i,{\bf k}^{'}_1,$
${\bf k}^{'}_{2},{\bf k}^{'}_3:                            
s\times i\times {\bf k}^{'}_1\times {\bf k}^{'}_2\times          
{\bf k}^{'}_3 \}$ as the space of possible outcomes (here $i$ labels
different states in the Bell basis ($i=1\div4$),
\begin{equation}
\mbox{\boldmath $\Psi$}^{(\pm)+}({\bf k}^{'}_2,{\bf k}^{'}_3)=
\frac{1}{\sqrt{2}\sqrt{2k^0_2}\sqrt{2k^0_3}}
\left(
{\bf A}^+({\bf k}^{'}_2,+){\bf A}^+({\bf k}^{'}_3,-)  \pm
{\bf A}^+({\bf k}^{'}_2,-){\bf A}^+({\bf k}^{'}_3,+)
\right),
\end{equation}
\begin{displaymath}
\mbox{\boldmath $\Phi$}^{(\pm)+}({\bf k}^{'}_2,{\bf k}^{'}_3)=
\frac{1}{\sqrt{2}\sqrt{2k^0_2}\sqrt{2k^0_3}}
\left(
{\bf A}^+({\bf k}^{'}_2,+){\bf A}^+({\bf k}^{'}_3,+)  \pm
{\bf A}^+({\bf k}^{'}_2,-){\bf A}^+({\bf k}^{'}_3,-)
\right).
\end{displaymath}
The identity resolution itself is
\begin{equation}
I_3=
\sum_{s=\pm}
\int\int\int
\frac{d {\bf k}^{'}_1}{\sqrt{2k_1^{0'}}}
\frac{d {\bf k}^{'}_2}{\sqrt{2k_2^{0'}}}
\frac{d {\bf k}^{'}_3}{\sqrt{2k_3^{0'}}}
\end{equation}
\begin{displaymath}
\left\{
{\bf A}^+({\bf k}^{'}_1,s)
\left(
\mbox{\boldmath $\Psi$}^{(+)+}({\bf k}_2^{'},{\bf k}_3^{'})|0\rangle
\langle 0|\mbox{\boldmath $\Psi$}^{(+)-}({\bf k}_2^{'},{\bf k}_3^{'})
+
\mbox{\boldmath $\Psi$}^{(-)+}({\bf k}_2^{'},{\bf k}_3^{'})|0\rangle
\langle 0|\mbox{\boldmath $\Psi$}^{(-)-}({\bf k}_2^{'},{\bf k}_3^{'})
+
\right.
\right.
\end{displaymath}
\begin{displaymath}
\left.
\left.
\mbox{\boldmath $\Phi$}^{(+)+}({\bf k}_2^{'},{\bf k}_3^{'})|0\rangle
\langle 0|\mbox{\boldmath $\Phi$}^{(+)-}({\bf k}_2^{'},{\bf k}_3^{'})
+
\mbox{\boldmath $\Phi$}^{(-)+}({\bf k}_2^{'},{\bf k}_3^{'})|0\rangle
\langle 0|\mbox{\boldmath $\Phi$}^{(-)-}({\bf k}_2^{'},{\bf k}_3^{'})
\right)
{\bf A}^-({\bf k}^{'}_1,s)
\right\}
\end{displaymath}
The initial state (an EPR-pair and a photon in the unknown polarization state)
is written as
\begin{equation}
|\mbox{\boldmath $\psi$}({\bf f})
\mbox{\boldmath $\Psi$}^{(-)}({\bf k}_2,{\bf k}_3)\rangle=
\mbox{\boldmath $\psi$}^{+}({\bf f})
\mbox{\boldmath $\Psi$}^{(-)+}({\bf k}_2,{\bf k}_3)|0\rangle=
\end{equation}
\begin{displaymath}
\left\{
-
\mbox{\boldmath $\Psi$}^{(-)+}({\bf k}_1,{\bf k}_2)
(
f(+) \mbox{\boldmath $\psi$}^+({\bf k}_3,+)+
f(-) \mbox{\boldmath $\psi$}^+({\bf k}_3,-)
)+
\right.
\end{displaymath}
\begin{displaymath}
\left.
\mbox{\boldmath $\Psi$}^{(+)+}({\bf k}_1,{\bf k}_2)
(
-f(+) \mbox{\boldmath $\psi$}^+({\bf k}_3,+)+
f(-) \mbox{\boldmath $\psi$}^+({\bf k}_3,-)
)+
\right.
\end{displaymath}
\begin{displaymath}
\left.
\mbox{\boldmath $\Phi$}^{(-)+}({\bf k}_1,{\bf k}_2)
(
f(+) \mbox{\boldmath $\psi$}^+({\bf k}_3,-)+
f(-) \mbox{\boldmath $\psi$}^+({\bf k}_3,+)
)+
\right.
\end{displaymath}
\begin{displaymath}
\left.
\mbox{\boldmath $\Phi$}^{(+)+}({\bf k}_1,{\bf k}_2)
(
f(+) \mbox{\boldmath $\psi$}^+({\bf k}_3,-)-
f(-) \mbox{\boldmath $\psi$}^+({\bf k}_3,+)
)
\right\}
|0\rangle.
\end{displaymath}
The fact that the momenta of all the particles are known allows to
preform the post-selection after the measurement, i.e. to discard all
the outcomes except for those in which the total momentum of the measured
particles ${\bf k}_{1}^{'}+{\bf k}_{2}^{'}={\bf k}_{2}+{\bf k}_{2}$,
and the teleported state is selected by the condition
${\bf k}_{3}^{'}={\bf k}_{3}$. In other words, kept in the two-particle
channel of the measurement are only the outcomes where the total momentum
is equal to the sum of the momentum of one of the particles of the
original EPR-pair and the momentum of the particle in the unknown
polarization state.

The probabilities of different outcomes after the post-selection
in the outcome space in the vicinity of
${\bf k}_{1}^{'}+{\bf k}_{2}^{'}={\bf k}_{1}+{\bf k}_{2}$ and
${\bf k}_{3}^{'}={\bf k}_{3}$) are determined by a formula
similar to Eqs. (23--24); for example, in the
channel associated with the projection on the state
$\mbox{\boldmath $\Psi$}^{(-)-}({\bf k}_1,{\bf k}_2)$,
we obtain for the amplitude
\begin{equation}
{\cal A}_{s}^{-}({\bf k}_1,{\bf k}_2,{\bf k}_3)=
\langle 0|
\mbox{\boldmath $\Psi$}^{(-)-}{\bf k}_1,{\bf k}_2)
\mbox{\boldmath $\psi$}^{-}({\bf k}_3,s)
\mbox{\boldmath $\psi$}^{+}({\bf f})
\mbox{\boldmath $\Psi$}^{(-)+}{\bf k}_2,{\bf k}_3)
|0\rangle=
\end{equation}
\begin{displaymath}
\frac{1}{2 (2k^0_12k^0_22k^0_3)} \cdot
\end{displaymath}
\begin{displaymath}
\left\{
\langle 0|-
\mbox{\boldmath $\Psi$}^{(-)-}({\bf k}_1,{\bf k}_2)
\mbox{\boldmath $\psi$}^{-}({\bf k}_3,s)
\mbox{\boldmath $\Psi$}^{(-)+}({\bf k}_1,{\bf k}_2)
\left(
f(+)\mbox{\boldmath $\psi$}^{+}({\bf k}_3,+)+
f(-)\mbox{\boldmath $\psi$}^{+}({\bf k}_3,-)
\right)
|0\rangle
+
\right.
\end{displaymath}
\begin{displaymath}
\langle 0|
\mbox{\boldmath $\Psi$}^{(-)-}({\bf k}_1,{\bf k}_2)
\mbox{\boldmath $\psi$}^{-}({\bf k}_3,s)
\mbox{\boldmath $\Psi$}^{(-)+}({\bf k}_1,{\bf k}_2)
\left(
-f(+)\mbox{\boldmath $\psi$}^{+}({\bf k}_3,+)+
f(-)\mbox{\boldmath $\psi$}^{+}({\bf k}_3,-)
\right)
|0\rangle
+
\end{displaymath}
\begin{displaymath}
\langle 0|
\mbox{\boldmath $\Psi$}^{(-)-}({\bf k}_1,{\bf k}_2)
\mbox{\boldmath $\psi$}^{-}({\bf k}_3,s)
\mbox{\boldmath $\Phi$}^{(-)+}({\bf k}_1,{\bf k}_2)
\left(
-f(+)\mbox{\boldmath $\psi$}^{+}({\bf k}_3,-)+
f(-)\mbox{\boldmath $\psi$}^{+}({\bf k}_3,+)
\right)
|0\rangle
+
\end{displaymath}
\begin{displaymath}
\left.
\langle 0|
\mbox{\boldmath $\Psi$}^{(-)-}({\bf k}_1,{\bf k}_2)
\mbox{\boldmath $\psi$}^{-}({\bf k}_3,s)
\mbox{\boldmath $\Phi$}^{(+)+}({\bf k}_1,{\bf k}_2)
\left(
f(+)\mbox{\boldmath $\psi$}^{+}({\bf k}_3,-)-
f(-)\mbox{\boldmath $\psi$}^{+}({\bf k}_3,+)
\right)
|0\rangle
\right\}.
\end{displaymath}
Finally for the propagation amplitude one obtains
\begin{equation}
{\cal A}_{s}^{-}({\bf k}_1,{\bf k}_2,{\bf k}_3)\propto
\langle 0|
{\bf A}^-(s,{\bf k}_3)
\left(
f(+){\bf A}^{+}({\bf k}_3,+)+
f(-){\bf A}^{+}({\bf k}_3,-)
\right)
|0\rangle,
\end{equation}
which actually coincides with the amplitude of propagation of
the state with the initial unknown polarization
along the direction defined by ${\bf k}_3$ instead of ${\bf k}_1$.
A similar situation takes place for different outcomes in the other
channels where the teleported state coincides to within a unitary
transformation completely determined by the channel number
(Bell basis vector number) with the unknown state which had
to be teleported.

Thus, if the state to be teleported is completely unknown, one can only
speak of the total amplitude of the propagation of the field as a whole.
On the other hand, if the state is only partly unknown, the available
{\it a priori} information can be used to ``label'' the particles to
make the identical particles effectively distinguishable.

This work was supported by the Russian Foundation for Basic Research
(project No 99-02-18127), the project ``Physical Principles of the
Quantum Computer'',  and the program ``Advanced Devices and Technologies
in Micro- and Nanoelectronics''.

This work was also supported by the Wihuri Foundation, Finland.

\end{document}